\documentstyle[prd,aps,12pt,epsf,psfig]{revtex}
\begin{document}

\baselineskip=7mm
\def\ap#1#2#3{           {\it Ann. Phys. (NY) }{\bf #1} () #3}
\def\arnps#1#2#3{        {\it Ann. Rev. Nucl. Part. Sci. }{\bf #1} () #3}
\def\cnpp#1#2#3{        {\it Comm. Nucl. Part. Phys. }{\bf #1} (#2) #3}
\def\apj#1#2#3{          {\it Astrophys. J. }{\bf #1} () #3}
\def\asr#1#2#3{          {\it Astrophys. Space Rev. }{\bf #1} (#2) #3}
\def\ass#1#2#3{          {\it Astrophys. Space Sci. }{\bf #1} (#2) #3}

\def\apjl#1#2#3{         {\it Astrophys. J. Lett. }{\bf #1} (#2) #3}
\def\ass#1#2#3{          {\it Astrophys. Space Sci. }{\bf #1} (#2) #3}
\def\jel#1#2#3{         {\it Journal Europhys. Lett. }{\bf #1} (#2) #3}

\def\ib#1#2#3{           {\it ibid. }{\bf #1} (#2) #3}
\def\nat#1#2#3{          {\it Nature }{\bf #1} (#2) #3}
\def\nps#1#2#3{          {\it Nucl. Phys. B (Proc. Suppl.) } {\bf #1} (#2) #3}
\def\np#1#2#3{           {\it Nucl. Phys. }{\bf #1} (#2) #3}

\def\pl#1#2#3{           {\it Phys. Lett. }{\bf #1} (#2) #3}
\def\pr#1#2#3{           {\it Phys. Rev. }{\bf #1} (#2) #3}
\def\prep#1#2#3{         {\it Phys. Rep. }{\bf #1} (#2) #3}
\def\prl#1#2#3{          {\it Phys. Rev. Lett. }{\bf #1} (#2) #3}
\def\pw#1#2#3{          {\it Particle World }{\bf #1} (#2) #3}
\def\ptp#1#2#3{          {\it Prog. Theor. Phys. }{\bf #1} (#2) #3}
\def\jppnp#1#2#3{         {\it J. Prog. Part. Nucl. Phys. }{\bf #1} (#2) #3}

\def\rpp#1#2#3{         {\it Rep. on Prog. in Phys. }{\bf #1} (#2) #3}
\def\ptps#1#2#3{         {\it Prog. Theor. Phys. Suppl. }{\bf #1} (#2) #3}
\def\rmp#1#2#3{          {\it Rev. Mod. Phys. }{\bf #1} (#2) #3}
\def\zp#1#2#3{           {\it Zeit. fur Physik }{\bf #1} (#2) #3}
\def\fp#1#2#3{           {\it Fortschr. Phys. }{\bf #1} (#2) #3}
\def\Zp#1#2#3{           {\it Z. Physik }{\bf #1} (#2) #3}
\def\Sci#1#2#3{          {\it Science }{\bf #1} (#2) #3}

\def\n.c.#1#2#3{         {\it Nuovo Cim. }{\bf #1} (#2) #3}
\def\r.n.c.#1#2#3{       {\it Riv. del Nuovo Cim. }{\bf #1} (#2) #3}
\def\sjnp#1#2#3{         {\it Sov. J. Nucl. Phys. }{\bf #1} (#2) #3}
\def\yf#1#2#3{           {\it Yad. Fiz. }{\bf #1} (#2) #3}
\def\zetf#1#2#3{         {\it Z. Eksp. Teor. Fiz. }{\bf #1} (#2) #3}
\def\zetfpr#1#2#3{         {\it Z. Eksp. Teor. Fiz. Pisma. Red. }{\bf #1} (#2) #3}
\def\jetp#1#2#3{         {\it JETP }{\bf #1} (#2) #3}
\def\mpl#1#2#3{          {\it Mod. Phys. Lett. }{\bf #1} (#2) #3}
\def\ufn#1#2#3{          {\it Usp. Fiz. Naut. }{\bf #1} (#2) #3}
\def\sp#1#2#3{           {\it Sov. Phys.-Usp.}{\bf #1} (#2) #3}
\def\ppnp#1#2#3{           {\it Prog. Part. Nucl. Phys. }{\bf #1} (#2) #3}
\def\cnpp#1#2#3{           {\it Comm. Nucl. Part. Phys. }{\bf #1} (#2) #3}
\def\ijmp#1#2#3{           {\it Int. J. Mod. Phys. }{\bf #1} (#2) #3}
\def\ic#1#2#3{           {\it Investigaci\'on y Ciencia }{\bf #1} (#2) #3}
\def\tp{these proceedings}
\def\pc{private communication}
\def\ip{in preparation}
\newcommand{\TeV}{\,{\rm TeV}}
\newcommand{\GeV}{\,{\rm GeV}}
\newcommand{\MeV}{\,{\rm MeV}}
\newcommand{\keV}{\,{\rm keV}}
\newcommand{\eV}{\,{\rm eV}}
\newcommand{\Tr}{{\rm Tr}\!}
\renewcommand{\arraystretch}{1.2}
\newcommand{\be}{\begin{equation}}
\newcommand{\ee}{\end{equation}}
\newcommand{\bea}{\begin{eqnarray}}
\newcommand{\eea}{\end{eqnarray}}
\newcommand{\ba}{\begin{array}}
\newcommand{\ea}{\end{array}}
\newcommand{\bmat}{\left(\ba}
\newcommand{\emat}{\ea\right)}
\newcommand{\refs}[1]{(\ref{#1})}
\newcommand{\ler}{\stackrel{\scriptstyle <}{\scriptstyle\sim}}
\newcommand{\ger}{\stackrel{\scriptstyle >}{\scriptstyle\sim}}
\newcommand{\lag}{\langle}
\newcommand{\rag}{\rangle}
\newcommand{\ns}{\normalsize}
\newcommand{\cm}{{\cal M}}
\newcommand{\gr}{m_{3/2}}
\newcommand{\p}{\partial}
\renewcommand{\le}{\left(}
\newcommand{\ri}{\right)}
\relax
\def\321{$SU(3)\times SU(2)\times U(1)$}
\def\21{$SU(2)\times U(1)$}
\def\dbd{$0\nu\beta\beta$ }
\def\emt{$L_e-L_\mu-L_\tau$ }
\def\mee{m_{ee}}
\def\ord{{\cal O}}
\def\tl{{\tilde{l}}}
\def\tL{{\tilde{L}}}
\def\bd{{\overline{d}}}
\def\tL{{\tilde{L}}}
\def\la{$m_{ee}~(\eV)$}
\def\a{\alpha}
\def\b{\beta}
\def\g{\gamma}
\def\c{\chi}
\def\d{\delta}
\def\D{\Delta}
\def\db{{\overline{\delta}}}
\def\Db{{\overline{\Delta}}}
\def\e{\epsilon}
\def\l{\lambda}
\def\n{\nu}
\def\m{\mu}
\def\nt{{\tilde{\nu}}}
\def\p{\phi}
\def\P{\Phi}
\def\solm{\Delta_{\odot}}
\def\sola{\theta_{\odot}}
\def\mee{m_{ee}}
\def\atm{\Delta_{\makebox{\tiny{\bf atm}}}}
\def\k{\kappa}
\def\x{\xi}
\def\r{\rho}
\def\s{\sigma}
\def\t{\tau}
\def\th{\theta}
\def\om{\omega}
\def\ne{\nu_e}
\def\nm{\nu_{\mu}}
\def\snui{\tilde{\nu_i}}
\def\ehat{\hat{e}}
\def\la{{\makebox{\tiny{\bf loop}}}}
\def\ta{\tilde{a}}
\def\tb{\tilde{b}}
\def\mb{m_{1b}}
\def\mt{m_{1 \tau}}
\def\rl{{\rho}_l}
\def\meg{\m \rightarrow e \g}
\def\mnu0{{\cal M}_{0\nu}}
\def\dt{\delta_{\tau}}
\renewcommand{\Huge}{\Large}
\renewcommand{\LARGE}{\Large}
\renewcommand{\Large}{\large}
\title{\large{\bf Predictive Framework with a Pair of }\\
\large{\bf Degenerate Neutrinos at a High Scale}}
\author{ Anjan S. Joshipura$^{1}$, Saurabh D. Rindani$^{1}$ and N. Nimai
Singh$^{2}$\\[.5cm]
{\ns\it  $^1$Theoretical Physics Group, Physical Research Laboratory,}\\
{\ns\it Navrangpura, Ahmedabad, 380 009, India}\\
{\ns\it $^2$ Dept. of Physics, Gauhati University, Guwahati, 781 014, 
India}}
\maketitle \vskip .5cm
\begin{center}
{\bf Abstract}
\end{center}
Radiative generation of the solar scale $\solm$ is discussed in
the presence of leptonic  CP violation. We assume that both
the solar scale and $U_{e3}$ are zero at a high scale and the weak
radiative corrections generate them. It is shown that all 
leptonic mass matrices satisfying these requirements lead to a
unique prediction $\solm \cos 2\sola\approx 4 \dt \sin^2 \theta_A
|m_{ee}|^2$ for the solar scale in terms of the radiative
correction parameter $\dt$, the physical solar (atmospheric) mixing
angles
$\sola (\theta_A)$ and the Majorana neutrino mass $m_{ee}$ probed
in neutrinoless double beta decay. This relation is independent
of the mixing matrix and CP-violating phases at the high scale. The
presence of CP-violating phases leads to dilution in the solar
mixing angle defined at the high scale. Because of this, bi-maximal
mixing pattern at the high energy leads to large but non-maximal
solar mixing in the low-energy theory. An illustrative model with
this feature is discussed.
\vskip.5cm
\newpage

\section{Introduction:} The neutrino masses and mixing
pattern implied \cite{revs,sf} by the neutrino oscillation
experiments requires ({\em i}) two large mixing angles $\sola$ and
$\theta_A$ to explain respectively the results of the solar and
the atmospheric neutrino experiments, ({\em ii}) corresponding mass
scales $\solm$ and $\atm$ satisfying
\be \label{ratio} {\solm\over \atm}\approx 3\times 10^{-2}  \ee
and ({\em iii}) a third mixing angle which is very small $\theta_3\leq
9^0$.

The smallness of $\theta_3$ compared to other two large angles and
that of ${\solm\over \atm}$ are two of the major puzzles
\cite{bar} in neutrino physics requiring theoretical explanation.
One possibility is to suppose that some symmetry leads to
vanishing of $\solm$ or $\theta_3$ or both and its breaking is
responsible for small values of these parameters. Vanishing of
$\solm$ can be a consequence of lepton like $U(1)$ or more general
non-abelian symmetry such as $SU(2)_H$. The vanishing of
$\theta_3$ can also be attributed to some symmetry, e.g., \emt
invariance \cite{emtsym} studied extensively in the
literature. One
needs to provide a mechanism for symmetry breaking in order to
generate the solar scale. An economical possibility is to suppose
that physics at a high scale leads to vanishing solar scale and  the
standard weak gauge bosons (and their superpartners in
supersymmetric theory) are responsible for its generation at a low
scale \cite{wolf,pdp,pred}. This provides a well-defined symmetry
breaking pattern which can be deduced \cite{radrev} by studying the
evolution of the neutrino mass matrix through the renormalization
group (RG) equations \cite{rg}. This evolution has been studied
very extensively \cite{rad} with a different physical motivation.

Let us suppose that neutrino masses $m_{\nu_i}$ and the mixing
angle $\theta_3$ generated by  physics at a high scale (e.g.,
seesaw mechanism) are given by
\be \label{pdmass} m_{\nu_i}=(m,-m, m') ~~~~~;~~~~~ \sin\theta_3\equiv
|(U_0)_{e3}|=0 ~, \ee
where $U_0$ is the neutrino mixing matrix  at the high scale. Clearly,
a very large class of the
neutrino and the charged lepton mass matrices can lead to
eq.(\ref{pdmass}). The RG evolution
provides a systematic way to study generation of the solar scale
and $U_{e3}$ at a low energy in all these models. It was found in
\cite{pap2} that all models of leptonic mass matrices leading
to eq.(\ref{pdmass}) give a unique prediction for the solar scale
when CP is conserved,
\be \label{pred} \solm \cos 2\sola=4 \delta_{\tau} s_A^2
|m_{ee}|^2 +\ord(\delta_\m,\delta_{\tau}^2)~. \ee
Here, $m_{ee}$ is the effective neutrino mass probed in neutrinoless double
beta decay (\dbd )
and $\delta_\tau$ denotes \cite{radrev} the size of the
radiative correction induced by the Yukawa coupling of the $\tau$:
\be \label{deltai}
\delta_\tau\approx c\left({m_\tau\over 4 \pi v }\right)^2 \ln{M_X\over M_Z}~.
\ee
$c=\frac{3}{2},-\frac{1}{\cos^2\b}$ in respective cases of the
standard model (SM) and the minimal sypersymmetric standard model (MSSM).

This equation implies a strong correlation between the solar scale
and $m_{ee}$. In particular, it requires that $m_{ee}$ should
be close to its present limit if the large mixing angle (LMA)
solution \cite{revs} is to be reproduced. CP conservation was
assumed in the analysis presented in \cite{pap2}. When CP
violation is allowed, the vanishing of $\solm$ at high scale
implies that the first two masses are degenerate in
eq.(\ref{pdmass}) only up to a phase. It is well-known
\cite{radrev,rad,haba} that CP violating phases can alter
evolution of the neutrino masses in a non-trivial and drastic
manner. It is thus important to include the effects of such
phases. We find that the CP violating phases $\a,\b$ associated
with the neutrino masses influence the predicted value of the
solar mixing angle $\sola$ significantly but in such a way that the basic
prediction eq.(\ref{pred}) obtained in a CP conserving theory
remains unaffected. Unlike in eq.(\ref{pred}), the radiatively
generated $U_{e3}$ depends on the phases $\a$ and $\b$. In the
following, we derive these basic results and  discuss their
consequences.
\section{Assumptions and Results}
A complex symmetric neutrino mass matrix can always be
diagonalized by a unitary  matrix $U$. This $U$  can be identified
with the MNS \cite{mns} matrix when the neutrino mass matrix is
specified in the flavour basis. $U$  is known to be determined by
three mixing angles $\theta_i$ and three phases $\d,\a,\b$ and can
be parameterized as
\be \label{ugen} U=R_{23}(\theta_{2})R_{13}(\theta_3 e^{-i
\delta})R_{12}(\theta_{1}) diag. (1, e^{{i \alpha \over 2}},e^{{i
\beta \over 2}}) \ee
The phase $\delta$ is analogous to the CKM phases and $\a,\b$ are
the phases associated with the Majorana masses.

We denote the neutrino mass matrix in the flavour basis at a high
scale by $\mnu0$ and the corresponding mixing matrix by $U_0$.
The $U_0$ is obtained by choosing $\theta_3=0$ in eq.(\ref{ugen}). 
Explicitly,
\be \label{u0}
U_0=\bmat{ccc} c_1&s_1&0\\-c_2 s_1&c_2 c_1&s_2\\s_1 s_2&-c_1 s_2&c_2 \emat
\bmat{ccc} 1&0&0\\0&e^{\frac{i\a}{2}}&0\\0&0&e^{\frac{i\b}{2}} \emat
~. \ee
where $c_1=\cos\theta_1,s_1=\sin{\theta_1}$ etc.

The solar scale vanishes if

\be \label{u0mnu} U_0^T \mnu0 U_0=diag.(m,m,m') \ee

Several points are to be noted in connection with the above
equation.
\begin{itemize}
\item One can choose $m$ and $m'$ real and positive without loss
of generality. The non-zero $|m'^2-m^2|$ can be identified with
the atmospheric scale $\atm$.
\item Given the mass ordering as in eq.(\ref{u0mnu}), $(U_0)_{e3}$
denotes the mixing element probed at CHOOZ. We assumed it to be
zero which amounted to choosing $\theta_{3}=0$ in eq.(\ref{ugen}).
This then implies that the CKM phase can be assumed zero at a high
scale. CP violation is still present through non-vanishing $\a$
and $\b$.
\item The neutrino mass matrix $\mnu0$ can be determined
by
inverting eq.(\ref{u0mnu}):
\be \label{zeroorder} \mnu0=U_0^* diag.(m,m,m')U_0^\dagger \ee
The matrix $U_0$ is however not unique. Degeneracy of masses in
eq.(\ref{u0mnu}) implies that $U_0$ is arbitrary up to an orthogonal
transformation $O_{12}$ by an angle $\theta_{1}'$ in the $12$
plane. Thus $U_0$ and $U_0 O_{12}$ imply the same physics. This
arbitrariness in $U_0$ amounts to the following redefinition of the angle
$\theta_{1}$ appearing in eq.(\ref{u0}):
\bea \label{arbi}|c_{1}|&\rightarrow& |c_{1}c_{1}'-s_{1} s_{1}'
e^{{i\a\over 2}}|~, \nonumber \\
|s_1|&\rightarrow& |c_{1}s_{1}'+s_{1} c_{1}' e^{{i\a\over 2}}| ~,
\eea
This freedom implies that the solar angle corresponding to the
$12$ mixing cannot be uniquely defined at a high scale.
Eq.(\ref{arbi}) at the same time does not allow  us to completely
rotate away the solar angle unless $\a=2 n \pi$. The arbitrariness
in defining the solar angle will be removed by the radiative
corrections. We can thus set $\theta'_{1}$ to zero without loss of
generality.
\end{itemize}

The matrix $\mnu0$ determined by eq.(\ref{zeroorder}) is modified
by radiative corrections. The radiatively corrected form of
$\mnu0$ follows from the relevant RG equations. We assume RG
equations corresponding to the SM or the MSSM. The modified
neutrino mass matrix is given \cite{radrev} in this case by
\be \label{corrected} \mnu0\rightarrow {\cal M}_\nu \approx I_g
I_t ~(I~ U^*_0 ~diag.(m,m,m_3)~U^\dagger_0 ~I~)\ee
where $I_{g,t}$ are calculable numbers depending on the gauge and
top quark Yukawa couplings. $I$ is a flavour dependent matrix
given by
$$ I\approx diag.(1+\delta_e,1+\delta_\mu,1+\delta_\tau) ~.$$
$\delta_{e,\mu}$ are obtained from  eq.(\ref{deltai}) by replacing
the tau mass by the electron and the muon masses respectively. The
physical neutrino masses and mixing are obtained by diagonalizing
the above matrix. We do this approximately by retaining the
contribution of the tau Yukawa coupling and by working to the
lowest order in $\dt$. The effect of radiative corrections is
best seen by going to the basis in which the neutrino mass matrix
$\mnu0$ defined at high scale is diagonal. This is done through a
rotation by the original $U_0$ on eq.(\ref{corrected})
\bea \label{rotated}  \tilde{{\cal M}}_\nu&\equiv& U_0^T {\cal
M}_\nu U_0 \nonumber \\ &\approx& m I_g I_t\bmat{ccc} 1+2\dt s_1^2
s_2^2&-2 \dt \cos{\a\over 2} c_1 s_1
s_2^2&\dt c_2 s_2 s_1 (e^{{i \beta \over 2}}+re^{{-i \beta \over 2}} )\\
-2 \dt \cos{\a\over 2} c_1 s_1 s_2^2&(1+2 \dt
c_1^2 s_2^2)&-\dt c_2 s_2 c_1(e^{-i \chi}+r e^{i\chi})\\
\dt c_2 s_2 s_1(e^{{i \beta \over 2}}+r e^{-{i \beta \over 2}})
&-\dt c_2 s_2 c_1 (e^{-i \chi}+r e^{i\chi})&r(1+2 \dt c_2^2)\\
\emat  \eea
where $r\equiv {m'\over m}$ and $\chi\equiv {(\alpha-\beta)\over
2}$.

Approximate diagonalization of the above matrix is
straightforward. We omit the details of this diagonalization but
mention the salient points.
\begin{itemize}
\item The structure of the upper $2\times 2$ block in eq.(\ref{rotated})
implies that the correction to the solar mixing angle arising from
its diagonalization  is not $\ord(\dt)$ but $\ord(1)$ unless
$\a=\pi$. This is a well-studied phenomenon \cite{radrev} which is
sometimes referred to as  radiative instability of the mixing
angle. This feature is related here to the arbitrariness (eq.(\ref{arbi})) 
in defining the mixing angle at high scale.
Perturbation present in eq.(\ref{rotated}) helps in removing this
ambiguity.

The upper $2\times 2$ block of eq.(\ref{rotated}) is diagonalized
by
\be \label{u12} U_{12}=\bmat{ccc} \tilde{c}_1& \tilde{s}_1
&0\\
-\tilde{s}_1&\tilde{c}_1&0\\
0&0&1\\ \emat ~,\ee
with \be \label{tilde}\tan 2\tilde{\theta_1}=-\cos {\alpha\over 2}
\tan 2\theta_1 ~, \ee
\item The (13) and (23) elements of the matrix 
$U_{12}^T \tilde{{\cal M}}_\nu U_{12}$
resulting after rotation through $U_{12}$
are $ \ord(m \dt)$ and the 33 element is approximately $m'$. The effect of these
off-diagonal elements is to induce additional mixing of the $\ord(\e)$
where $\e\approx\frac{m\dt}{m'-m}\approx \frac{2 m^2\dt}{\atm}$. This mixing 
is quite small for physically interesting mass range $m\leq \ord(\eV)$.
Thus a complete diagonalization of $U_{12}^T \tilde{{\cal M}}_\nu U_{12}$
is performed by two additional rotations with mixing angles which are
$\ord(\e)$. Explicitly, 
\be \label{final} U_{23}^T U_{13}^T U_{12}^T U_0^T {\cal M}_\nu U_0 U_{12}
U_{13}U_{23}\approx diag.(m_{\nu_1},m_{\nu_2},m_{\nu_3}) ~. \ee
The $U_{13}$ is responsible for generation of the CHOOZ angle
while $U_{23}$ provides a small radiative correction to the 
atmospheric mixing angle $\theta_2$. We neglect correction to $\theta_2$ in
the
following. $U_{13}$ is given by
\be \label{u13} U_{13}=\bmat{ccc} \tilde{c}_3&0& \tilde{s}_3
e^{i\Psi}\\
0&1&0\\
-\tilde{s}_3 e^{-i\Psi}&0&\tilde{c}_3\\ \emat ~,\ee
\end{itemize}

Eq.(\ref{final}) gives us the complete mixing matrix which can be
approximately written as
\bea \label{vint}  U&\approx &U_0U_{12}U_{13} \nonumber \\
& \approx &\bmat{ccc}
c_{\odot}\tilde{c}_3 e^{-i \eta}&i s_{\odot} e^{-i \eta}& c_{\odot}
\tilde{s}_3 e^{i(\Psi-\eta)}\\
i c_A s_{\odot}\tilde{c}_3 e^{i (\eta+\a/2)}-s_A\tilde{s}_3
e^{(i\beta/2-\psi)}&c_A c_{\odot}e^{i(\eta+\a/2)}&
ic_A s_{\odot}\tilde{s}_3 e^{i (\psi+\eta+\a/2)}+s_A
\tilde{c}_3 e^{i\beta/2}\\
-i s_A s_{\odot}\tilde{c}_3 e^{i (\eta+\a/2)}-c_A\tilde{s}_3
e^{i(\beta/2-\psi)}&-s_Ac_{\odot}e^{i(\eta+\a/2)}&-i s_A
s_{\odot}\tilde{s}_3 e^{i\psi }+c_A\tilde{c}_3 e^{i\beta/2}\\ \emat ~.\eea
where,
\bea \label{angs} c_{\odot}e^{i \eta}&=& (c_1 \tilde{c}_1-s_1
\tilde{s}_1 e^{-i {\a\over 2}}) ~,\nonumber \\
s_\odot e^{i \eta}&=& i(c_1 \tilde{s}_1+s_1
\tilde{c}_1 e^{-i{\a\over 2}}) ~,\nonumber \\
s_A&=&s_2 + \ord(\dt) ~. \eea
$\theta_3$ is radiatively generated and is given by
\bea \label{ue3} |U_{e3}|&=&|s_3|=|\tilde{s_3} c_{\odot}|=| \dt
s_A c_A c_\odot s_\odot {1+r \over 1-r} \cos\delta| \nonumber \\
&\approx& |\dt \sin 2 \theta_A \sin 2 \sola {m^2\over 2 \atm}
\cos\delta|~. \eea
where $r\equiv {m'\over m}$ can be expressed as:
\be \label{r}|r|=(1\pm {\atm\over m^2})^{1/2}\approx 1\pm
{\atm\over 2 m^2} \ee
The positive (negative) sign in the above equation applies to the 
case $m'>m$ ($m'<m$).
All the phases appearing in eq.(\ref{vfinal}) are expressible in
terms of the original phases $\alpha$ and $\beta$. Explicitly,
\bea \label{phases} \sin \eta&=&{s_1\over \sqrt{2} c_{\odot}} \sin
{\alpha\over 2}(1-{\cos 2 \theta_1 \over \cos 2 \sola})^{1/2} ~, \\
\beta_L&=&\pi+\beta-\alpha-2 \eta ~,\\
\delta&=&-{\beta_L\over 2} +\ord \left({\atm\over 2 m^2}\right)^2
~. \eea

The mixing matrix in eq.(\ref{vint}) assumes the following simple form
once non-leading terms
of $\ord(\tilde{s}_3)$ are neglected \footnote{In writing this form, we
have made a phase rotation on eq.(\ref{vint}) from the
left and right by two appropriate diagonal phase matrices $P_{L,R}$. The
$P_L$ is absorbed in redefining the charged lepton fields giving us the
final eq.(\ref{vfinal}).}:
\be \label{vfinal} U\equiv U_0U_{12}U_{13}\approx\bmat{ccc} c_{\odot} &s_{\odot}&s_3 e^{-i\delta}\\
                              -c_A s_{\odot}& c_A c_{\odot}&s_A \\
                              s_A s_{\odot}&-s_A c_{\odot}&c_A\\
                              \emat \bmat{ccc} 1&0&0\\
0&i&0\\0&0&e^{{i\beta_L\over 2}}\\ \emat ~. \ee

It is clear from eqs.(\ref{tilde},\ref{angs}) that the 
radiative corrections have removed the
ambiguity in the choice of the solar angle by fixing the
$\theta_1'$ in eq.(\ref{arbi}) to $\tilde{\theta}_1$ given in
eq.(\ref{tilde}). The same equation also determines the phase
$\eta$ as given in eq.(\ref{phases}). Interestingly enough, the
phase of the second mass state (which was $\alpha$ at the high
scale) is now determined to be ${\pi}$ in
eq.(\ref{vfinal}). This corresponds to (almost) equal and opposite
neutrino masses at the low scale.

The major physical effects of the radiative corrections are
generation of the solar scale, mixing angle $\theta_{3}$ and the
CKM phase $\delta$ which was absent at the high scale. The solar
scale follows from the eigenvalues of eq.(\ref{rotated}):
\bea \label{ev} m_{\nu_1}&\approx& I_g I_t m(1+2 \dt s_A^2
s_{\odot}^2)+\ord(\dt^2)
~,\nonumber \\
m_{\nu_2}&\approx&I_gI_t m (1+2 \dt s_A^2 c_{\odot}^2)+\ord(\dt^2)
~,\nonumber \\
m_{\nu_3}&\approx& I_gI_t m' (1+2 \dt c_A^2)+\ord(\dt^2) ~. \eea
which lead to
\be \label{sscale} \solm\equiv m_{\nu_2}^2-m_{\nu_1}^2=4
I_g^2I_t^2 m^2 \dt s_A^2 \cos 2\sola+\ord(\dt^2)~. \ee

The Majorana mass $m_{ee}$ is obtained using
eqs.(\ref{vfinal},\ref{ev}):
\be \label{bmass} |m_{ee}|^2\equiv |U_{ei}^2 m_{\nu_i}|^2\approx I_g^2
I_t^2 m^2 \cos^2 2 \sola
+\ord(\dt)~, \ee
The initial phases $\alpha,\beta$ appear in the above equation
only implicitly through the solar angle $\sola$. This is a
consequence of the fact that physical neutrinos resulting after
the radiative corrections consist of a pseudo-Dirac pair with
(almost) equal and opposite masses.

Eqs.(\ref{ue3},\ref{sscale}) are predictions of the scheme. The
common mass $m~I_g I_t $ of the degenerate pair can be identified
with the electron neutrino mass $m_{\nu_e}$ probed through the
direct neutrino mass search, e.g., in tritium beta decay \cite{mainz}. It
is
also probed through measurement of the Majorana mass parameter
$m_{ee}$ \cite{limits}. One can in fact eliminate $m$ from
eq.(\ref{sscale})
using eq.(\ref{bmass}). This leads to the prediction (\ref{pred})
already mentioned in the introduction. This prediction involves
only low energy measurable parameters and is independent of the CP
violating phase.
\section{Phenomenological Consequences}
The five observables, namely $\sola$, $\solm$, $|m_{ee}|$,
$m_{\nu_e}=m I_gI_t$ and $U_{e3}$, are correlated through
eqs.(\ref{pred},\ref{ue3},\ref{sscale}). We now study the
consequences of this correlation. $\dt$ is negative in case of
the MSSM. This implies a negative $\solm \cos2 \sola$ and hence
only much less preferred dark region of the solar parameter space.
This excludes the LMA and LOW solutions in case of the MSSM
but the SM can easily allow them. These solutions are realized
only for the specific range in $|m_{ee}|$. This is displayed in
Fig.(1) where we show contours of the $\solm$ and the electron
neutrino mass $m_{\nu_e}$ in the $\tan^2\sola$-$|m_{ee}|$ plane.
The values of $\tan^2\sola$ are restricted to the typical range
$\sim 0.2-0.8$ allowed by the LMA or LOW solution. For these
values, one obtains a $\solm$ in the required range
$10^{-4}-10^{-5}\eV^2$ provided $m_{ee}\sim 0.1-1 \eV$. This
value is close to the experimental limit \cite{limits}. We also
show the contours corresponding to the electron neutrino mass
$m_{\nu_e}$ equal to $0.5 \eV  $ and $2.0\eV$ in the same plot. It
is seen that $m_{\nu_e}$ is restricted to lie in the range $0.5-2
\eV$ in case of the LMA solution.

The predicted values of the $|U_{e3}|$ are also shown in the
figure. $|U_{e3}|$ is seen to be restricted to a typical range
$\sim 0.001-0.02$ in the region of the $m_{ee}-\tan^2\sola$ plane
allowed by the LMA solution.

The results presented above are completely in terms of the low
energy variables and do not need any knowledge of the parameters
at the high scale. Let us now comment on possible choices of the high
scale parameter. Eqs.(\ref{tilde},\ref{angs}) are equivalent to the
relation
\be \label{ths} \sin^2 2 \sola=\sin^2 2 \theta_1 \sin^2 {\alpha
\over 2} ~, \ee
The initial value of the solar mixing angle $\theta_1$ is subject
to the arbitrariness noted in eq.(\ref{arbi}). We see that
irrespective of this, any choice of $\theta_1$ and $\alpha$ must
satisfy
\be \label{th1alp} (\sin^2 2 \theta_1,\sin^2{\alpha\over 2})\geq
\sin^2 2\sola\sim  (0.6-0.9) ~. \ee
It is seen that the mixing angle is reduced compared to its value
at a high scale. This is phenomenologically interesting. One of
the preferred phenomenological schemes corresponds to 
bi-maximal mixing \cite{bim} which can arise from symmetry
considerations. However, the present solar data do not favour
strictly maximal mixing \cite{revs}. Eq.(\ref{ths}) shows that one
can start with bi-maximal mixing at a high scale and 
radiative corrections would lead to the desired reduction provided
the CP-violating phase $\alpha$ is chosen non-zero and different
from $\pi$ at a high scale. The presence of $\alpha$ also plays
another important role. $\a=\pi$ and maximal solar mixing
corresponds to vanishing $m_{ee}$ with the consequence that the
solar scale arise only at $\ord(\dt^2,\delta_\mu$), see
eq.(\ref{pred}). The natural value for the solar scale lies in the
vacuum region in this case \cite{barb,pap3}. An $\a$ different
from $\pi$ alters this and allows strictly  bi-maximal mixing 
and a non-zero $m_{ee}$ at a high scale.

The radiative reduction in the solar angle found here is to be
contrasted with a similar analysis presented recently in \cite{miura}.
This analysis assumed vanishing $U_{e3}$ but a non-zero $\solm$ at the
high scale
itself. It was then found \cite{miura} that radiative corrections tend to
increase the $\sin^2 2\sola$ compared to its value at the high scale.
This does not allow bi-maximal mixing at the high scale in contrast to
what is found here.

The analysis presented so far holds for all models satisfying
eq.(\ref{pdmass}) at a high scale. While many possibilities exist,
let us give an illustrative example which corresponds to 
bi-maximal mixing. This is specified by the following neutrino
mass matrix in the flavour basis.
\be \label{examp} \mnu0=R_{23}(\theta_2) \bmat{ccc} a&i b&0\\i
b&a&0\\0&0&m'e^{-i\b}\\ \emat R_{23}^T(\theta_2)~, \ee
The parameters $a$, $b$, $m'$ and $\theta_2$  are assumed real.

Define the mixing matrix $\tilde{U}_0$ as:
\be
\tilde{U}_0=e^{-{i\alpha \over 4}}R_{23}(\theta_2) \bmat{ccc}

{1\over \sqrt{2}}&{1\over \sqrt{2}}e^{{i\alpha \over 2}}&0\\
-{1\over \sqrt{2}}&{1\over \sqrt{2}}e^{{i\alpha \over 2}}&0\\
0&0&e^{i{\beta\over 2}}\\ \emat .\ee
This matrix diagonalizes eq.(\ref{examp}):

$$\tilde{U}_0^T~\mnu0~\tilde{U}_0=Diag. (m,m,m')$$,
with 
\bea \label{alphaex} \tan{\alpha\over 2}&=&-{b\over a}~,
\nonumber
\\
m^2&=&(a^2+b^2) ~.\eea
We thus have maximally mixed degenerate neutrinos at a high scale.
The maximal $\theta_1$ implies maximal $\tilde{\theta}_1$ through
eq.(\ref{tilde}). The solar angle $\sola$ and the low scale CP
violating phase follows respectively from eq.(\ref{angs}) and
eq.(\ref{phases}):
$$ \tan\sola=\cot{{\alpha\over 4}}~~~;~~~ \delta={\pi+\alpha-2 \beta\over
4} $$

It is seen that $\alpha\sim \pi$ leads to large solar mixing.
Specifically, $b\sim \sqrt{3} a$ leads to $\sola\sim 30^0$. The
radiatively generated solar scale can naturally fall in the LMA
region as already discussed before in the general context.
Eq.(\ref{examp}) in this way leads to the required pattern 
of neutrino masses and mixing.

The texture presented here is quite similar to the one studied in
ref \cite{barb} which assumed zero $a$ and a real $i b$.
For the reasons already mentioned, this model leads only to the
vacuum solution after radiative corrections are included. The
presence of $a$  and additional phase in $b$ alters this and
allows one to obtain the LMA solution.

It is possible \cite{pap3} to obtain the above texture in the context of
seesaw
model by invoking additional horizontal \footnote{$SU(2)_H$ symmetry 
has also been recently used
in \cite{mk} to generate bimaximal mixing pattern.} symmetry.
One way \cite{bar}
of realizing the above texture is to assume a charged lepton
mixing matrix with only $\mu-\tau$ mixing. This would generate
$R_{23}$ in eq.(\ref{examp}). The neutrino mass matrix in the weak
basis is then given by the block diagonal form explicitly
displayed in eq.(\ref{examp}). An explicit model was presented
with these features in \cite{pap3} in a CP conserving
situation. We do not elaborate on it here since a trivial
modification \footnote{This modification amounts 
to assuming a non-zero vacuum expectation value (vev) for the CP-odd field
$T^2$ and vanishing vev for the CP even field $T^1$ in the notation of
\cite{pap3}.} of 
this model incorporating CP 
violation leads to the mass matrix presented in eq.(\ref{examp}).
\section{Summary} The presently available information on neutrino
oscillations can be nicely understood if two of the neutrinos pair 
up to form a pseudo-Dirac state. This can be obtained from 
a degenerate neutrino pair by means of 
standard radiative corrections. We discussed basic predictions
of this picture including the important effects of the  CP-violating
phases. The scheme presented here has testable predictions: The
LMA solution requires $m_{ee}\sim 0.1- 1\eV$ close to its present limit,
relatively small $U_{e3}\sim 0.001-.01$ and observable $\sim 0.5
-2 \eV$ neutrino mass in beta decay. 

The CP violating phases $\a$ and $\b$ associated with the Majorana masses
do not effect the basic prediction (\ref{pred}) of the scheme but play
an important role in diluting the solar mixing angle defined at a high
scale. This allows bi-maximal mixing at the high scale and large but
non-maximal solar angle at the low scale in accordance with the demand of
the current solar neutrino results. \\ \\
{\bf Acknowledgments} Nimai Singh wishes to acknowledge local hospitality
at Physical Research Laboratory during initial part of this work.

\newpage
\begin{figure}[h]
\centerline{\psfig{figure=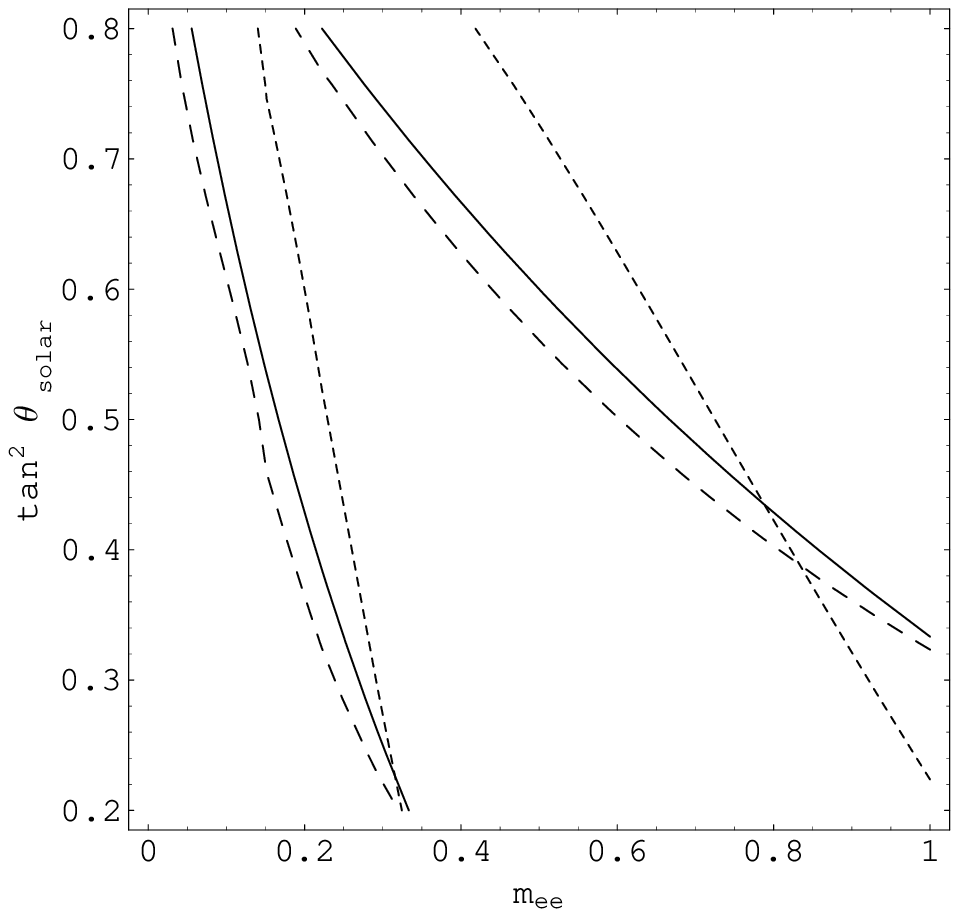,height=15cm,width=15cm}}
\vskip.25cm \caption{Contours of $\solm$ (dotted),
$|U_{e3}|$ (dashed) and the electron neutrino mass $m_{\nu_e}$
(solid) as a function of  $\tan^2\sola$ and $m_{ee}$ (in eV). The upper
(lower) curves
correspond to $\solm=10^{-4}\; (10^{-5})\; \eV^2,\; |U_{e3}|=0.02 \;
(0.01)$ and the $m_{\nu_e}=2.0 \;(0.5)\; \eV$ respectively.}
\end{figure}
\end{document}